# Orientation-Dependent Protein Binding at Nanoparticle Interfaces

*Vigneshwari Karunakaran Annapoorani*[1,2], *Ian Rouse*[1,2],

*Vladimir Lobaskin*[1,2], *Nicolae-Viorel Buchete*[1,2] *

[1]School of Physics, University College Dublin, Dublin 4, Belfield, Ireland
[2]Institute for Discovery, University College Dublin, Dublin 4, Belfield, Ireland

* Correspondence: buchete@ucd.ie

## ABSTRACT

Accurate quantification of protein–nanoparticle interactions is essential for applications in nanobiotechnology, nanomedicine, and drug delivery. Motivated by recent computational and experimental work, we combine coarse-grained united-atom (UA) models with molecular docking to characterize protein adsorption on $SiO_2$ nanoparticles. We construct orientation-resolved heatmaps in which polar and azimuthal angles uniquely specify the relative protein–nanoparticle pose, and the map amplitude reports binding propensity via the minimum UA adsorption energy or the docking score. Each angular bin corresponds to a distinct docked complex, enabling systematic comparison of binding geometries across models.

To relate docking score landscapes to Boltzmann-averaged UA adsorption energetics, we analyze eight birch pollen allergen proteins previously studied experimentally. Similarity between the two orientational distributions is quantified using the Jensen–Shannon divergence (JSD). We find encouraging agreement between the two approaches in several cases, while also identifying limitations and routes for improvement, including optimized angular resolution and iterative refinement of interaction parameters. Overall, this framework provides a quantitative bridge between coarse-grained energetics and docking outputs at protein–nanoparticle interfaces, supporting improved predictive modeling and mechanistic insight into protein–nanoparticle binding landscapes.

Studies of protein adsorption onto nanoscale materials date back to the early 1960s, initially motivated by liposomes and polymeric nanoparticles and investigated in both *ex vivo* and *in vivo* settings. In the late 2000s, the introduction of the "protein corona" concept[1] stimulated extensive work on lipid-, carbon-, and metal-based nanomaterials, often emphasizing *ex vivo* characterization. Subsequent efforts focused on the kinetics of corona formation, and particularly the long-lived "hard" corona.[1,2] By 2017, the broader notion of a "biomolecular corona" further expanded the scope to additional biological components and reinforced the need for integrated *ex vivo* and *in vivo* studies, with direct relevance to targeting, diagnosis, drug delivery, and other therapeutic applications.[3-7]

Engineered nanoparticles (NPs) are now widely explored for biomedical and therapeutic use, including drug delivery and molecular imaging, making a molecular-level understanding of bio–nano interactions a continuing priority.[4,7-10] Silica ($SiO_2$) NPs, in particular, are prevalent across medical, agricultural, and environmental contexts, motivating careful assessment of exposure and interfacial binding mechanisms. To address these challenges, computational approaches spanning hard-sphere and coarse-grained descriptions through molecular docking are increasingly used to probe the thermodynamics and structural evolution of NP–protein complexes.[11-18] A major goal is "corona fingerprinting,"[19] which seeks predictive links between NP–protein physicochemical features and downstream biological fate, including cellular uptake.[19] More broadly, combining computational modeling with experimental "green" synthesis and multi-spectroscopic characterization is helping to rationalize how intrinsic material properties govern emergent biological interactions and to inform nanosafety assessment.[3,5,14,20-27]

Recent computational studies of NP–biomolecule interactions have increasingly combined coarse-grained descriptions with atomistically informed tools, including the united-atom model (UAM),[16,17,28] and molecular docking approaches such as PatchDock.[29] For example, docking has been applied to probe corona formation on CuO and $TiO_2$ nanoparticles.[30] Despite this progress, experimental data that directly resolve corona formation pathways and interfacial properties remain limited, and improving quantitative corona analysis continues to be an active need.[31]

Whereas protein–protein interactions have been investigated systematically for decades,[32] coarse-grained frameworks tailored to protein–NP interactions (PNIs) have emerged more recently.[33] A key conclusion from these efforts is that — analogous to protein–protein interactions — PNIs depend strongly on relative orientation, not solely on intermolecular separation. Atomistically detailed methods would therefore be desirable; however, fully atomistic molecular dynamics is often prohibitively expensive for systematic sampling of adsorption orientations, even with high-performance computing. In practice, this makes coarse-grained models (e.g., UAM) and molecular docking among the few feasible routes for efficient, orientation-aware exploration of PNIs.

Building on recent work that highlighted both strengths and limitations of UAM-based PNI predictions, we employ molecular docking using PatchDock, one of the few docking platforms that explicitly supports NP–protein complexes. We have also shown that PatchDock can be used to compute statistical biophysical descriptors of NP–biomolecule complexes—such as hydrophobic and charged fractions of solvent-accessible surface area (SASA)—for proteins, oligomers, and more complex NP–protein corona structures.[34] However, docking score landscapes and coarse-grained adsorption energetics are rarely compared on a consistent quantitative footing.

Here, we introduce orientation-resolved PNI heatmaps and use the Jensen–Shannon divergence to directly compare docking-derived landscapes with UAM Boltzmann-averaged adsorption energies for eight allergen proteins on $SiO_2$ nanoparticles. This framework connects heatmap features to representative adsorption geometries, clarifies when docking can serve as a rapid proxy, and provides a route to refine both docking protocols and coarse-grained interaction models.

The UAM employed here was formulated and validated in recent publications.[35,36]
Here, the term UAM refers specifically to the coarse-grained protein model introduced by Lobaskin et al.[35,36] for the quantitative assessment of protein–nanoparticle interactions. It should not be confused with other "united-atom" approaches - such as those used in the GROMOS force field - in which hydrogen atoms are treated implicitly and merged with the adjacent carbon atoms. In prior work, Lobaskin and co-workers combined UAM-based *in silico* predictions with *in vitro* measurements of corona formation, including binding and selectivity across eight proteins on bare and coated $SiO_2$ NPs.[20] Following that benchmark set, we analyze the birch pollen–related proteins β-lactoglobulin A (PDB ID: 1CJ5), Bet v 2 (1CQA), lysozyme (1DPX), Bet v 4 (1H4B), ovalbumin (1UHG), Bet v 6 (2AS8), Bet v 1 (4A88), and serotransferrin (6JAS), which are shown in Figure 1.

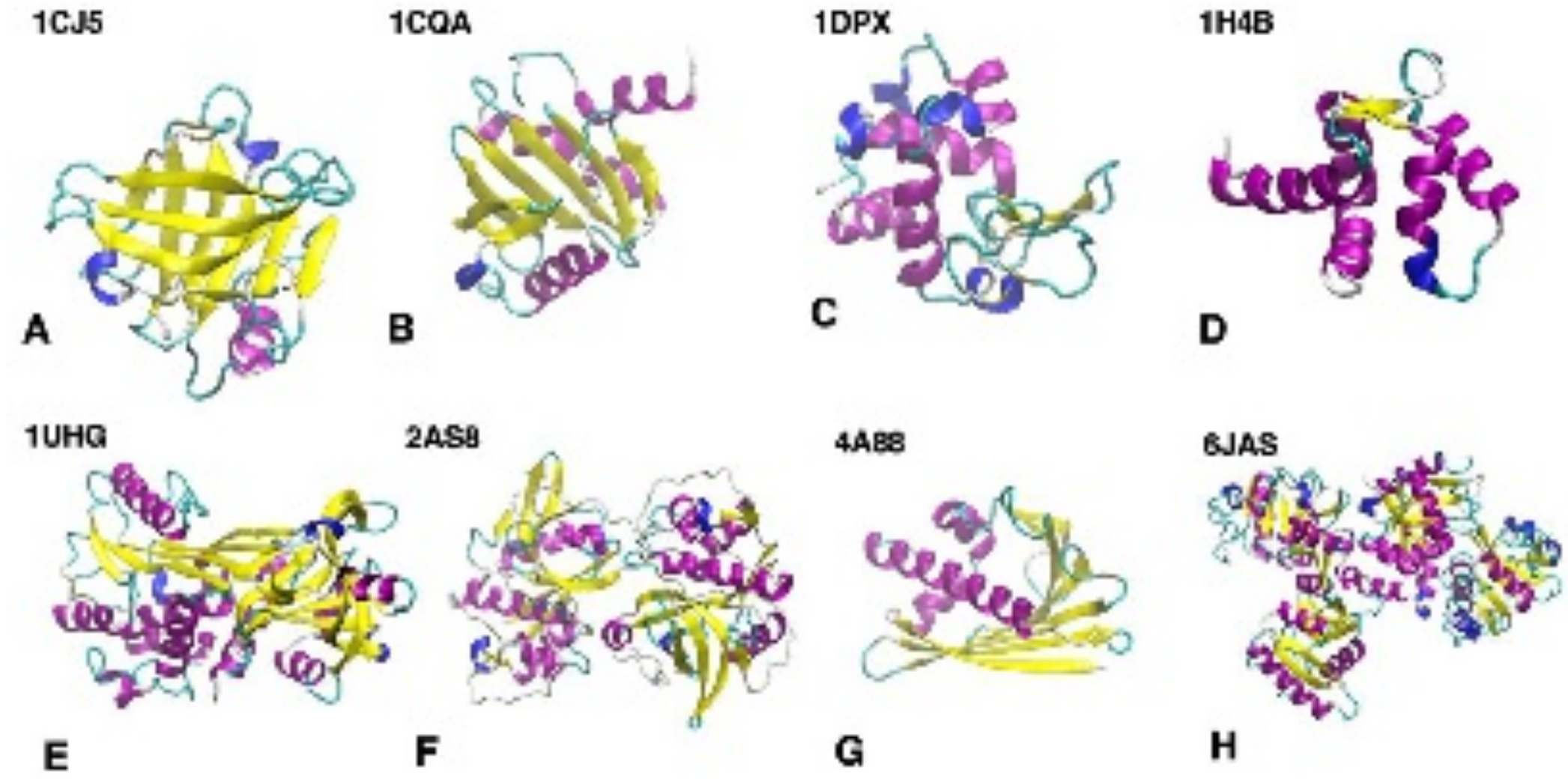

**Figure 1. Protein structures analyzed in this study**. Eight birch pollen–related proteins are shown to illustrate their size and conformational diversity: (A) β-lactoglobulin A (PDB ID: 1CJ5), (B) Bet v 2 (1CQA), (C) lysozyme (1DPX), (D) Bet v 4 (1H4B), (E) ovalbumin (1UHG), (F) Bet v 6 (2AS8), (G) Bet v 1 (4A88), and (H) serotransferrin (6JAS).

To clarify the reference frame and the procedure used to construct the heatmaps, we followed the UAM heatmap methodology described by Lopez et al. for the coarse-grained united-atom model.[36] The reference-frame definition is summarized in Figure 2.

As shown in Fig. 2(A), the protein orientation can be specified by selecting a point on the protein surface and defining the position vector from the protein center of mass (COM) to that point. The direction of this vector is parameterized by two angles, $\phi$ and $\theta$, in a spherical coordinate representation. A third coordinate, the COM separation, $d_{\mathrm{COM}}$, sets the distance along this direction. Depending on the nanomaterial geometry, $d_{\mathrm{COM}}$ is defined either relative to (B) the surface normal of a slab (protein–slab interaction, PSI) or (C) the center of a NP (protein-nanoparticle interaction, PNI).

Figure 2(D) illustrates the corresponding setup used for molecular docking. The NP center is identified, and the relative coordinates are expressed in spherical coordinates; the protein is treated as the receptor and the NP as the ligand (optionally represented by a selected subset of NP atoms).

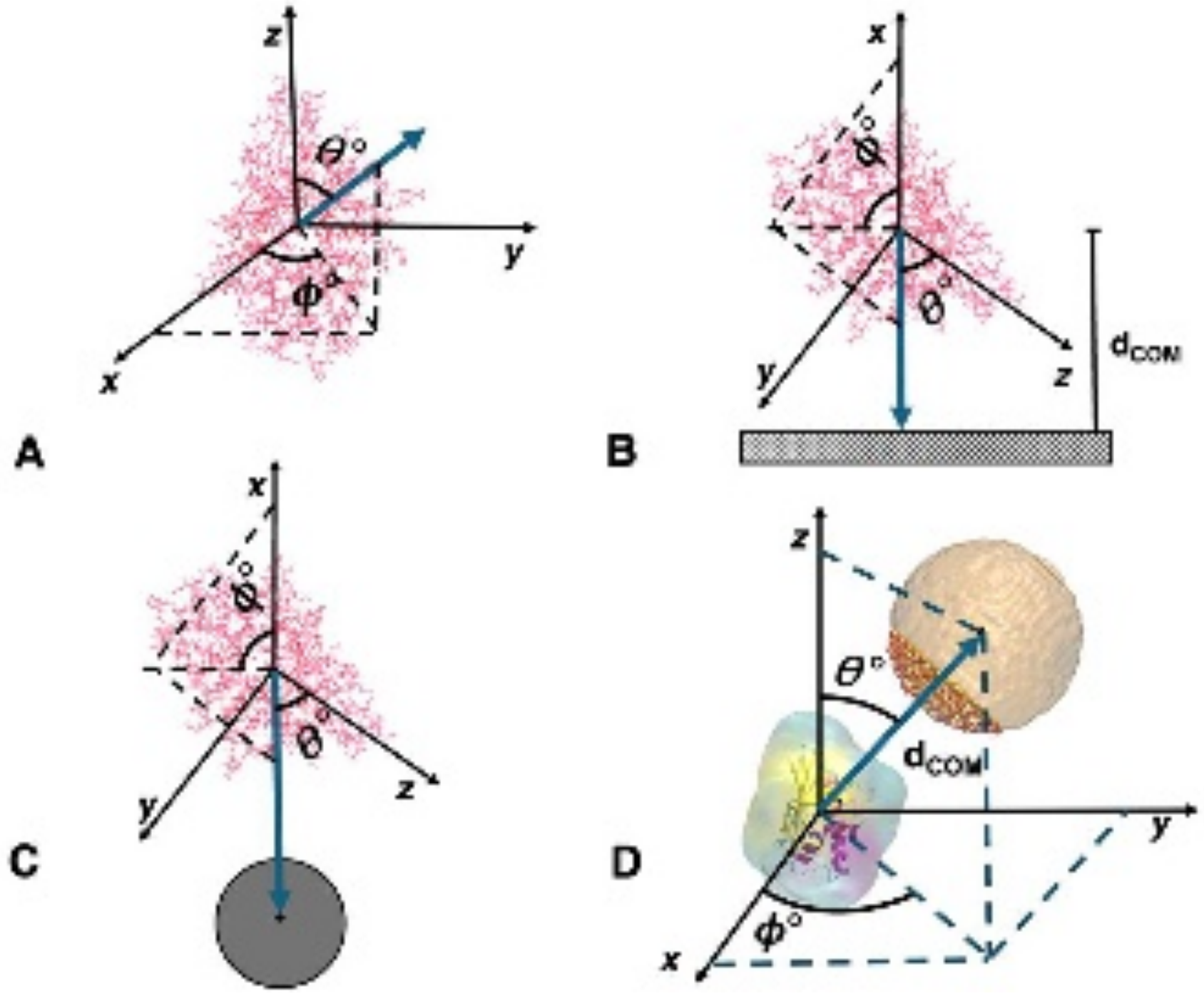

**Figure 2. Definition of protein–nanomaterial orientation.** (A) A protein-fixed spherical coordinate system is constructed from Cα Cartesian coordinates, with the origin at the protein center of mass (COMα) and axes aligned with the principal inertia axes (longest axis = x). Surface positions are specified by the angles $\theta$ and $\phi$. The remaining degree of freedom is the separation $d_{COM}$, measured relative to (B) the surface normal for a slab or (C) the nanoparticle center. (D) For spherical nanoparticles, each relative protein–NP pose (e.g., a docking result) is uniquely defined by the ($d_{COM}$, $\theta$, $\phi$) tuple. When the protein and NP are in atomic contact, the distance is constrained by the protein surface, and the angles $\theta$, $\phi$ alone uniquely specify the interaction pose used to generate PNI maps.

We note two practical limitations of docking: the search samples only a subset of possible complexes, and many docking implementations cannot treat an entire NP explicitly. In such cases, a representative slab segment of the NP surface can be used to restrict the search space.

Together, these definitions provide a consistent and compact description of protein pose and separation, enabling direct construction and comparison of orientation-resolved interaction heatmaps. We note that when the NP and the protein are in direct atomic contact (e.g., docked) the separation distance is constrained by the protein surface the angles $\theta$ and $\phi$ are sufficient to define uniquely the interaction pose, and can be used to generate protein-nanoparticle interaction (PNI) maps, as illustrated below.

As a first step, we performed UAM calculations using quartz silica parameters for the NP, a surface potential of −29 mV, and a radius of 5 nm. From these simulations we obtained orientation-specific adsorption energies and constructed ($\theta$, $\phi$) heatmaps, following the computational–experimental workflow previously used for NP–protein corona prediction and analysis and emphasizing FAIR data practices.[20]

In a second step, we carried out molecular docking as follows. (1) Protein preparation: each protein structure was prepared in a suitable format for docking (including, where appropriate, addition of hydrogen atoms). (2) NP model: a 10 nm-diameter $SiO_2$ NP model was generated (e.g., using CHARMM-GUI) by specifying the relevant composition and geometric parameters.[37] (3) Docking: docking between the NP and each protein was performed with PatchDock. To preserve the NP geometry, the $SiO_2$ particle was treated as a rigid body, enabling systematic sampling of protein orientations relative to the NP surface.[38] (4) Coordinate transformation: for each docked complex, the NP center was determined and the relative orientation was expressed via the spherical angles $\theta$ and $\phi$ by transforming Cartesian coordinates to spherical coordinates. (5) Angle assignment and scoring: ($\theta$, $\phi$) values were computed for all docking poses across the eight proteins. (6) Heatmap construction: PNI heatmaps were generated by mapping docking scores onto ($\theta$, $\phi$) bins, allowing preferred orientations to be identified and compared across proteins.

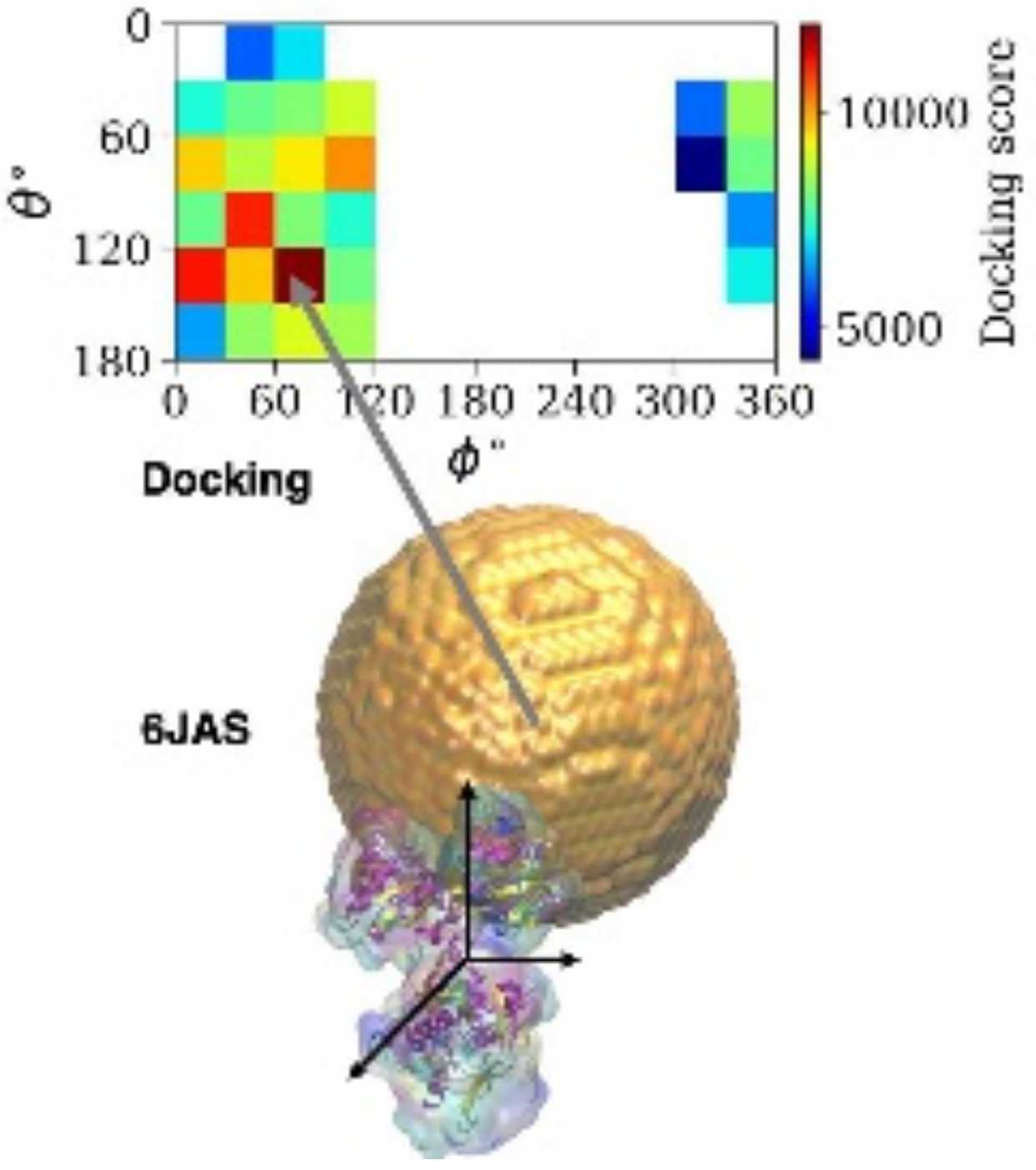

**Figure 3. Docking-derived protein–nanoparticle interaction (PNI) map for serotransferrin–$SiO_2$.** The PNI heatmap shows docking scores for serotransferrin (PDB ID: 6JAS) on a bare $SiO_2$ nanoparticle, with the highest-scoring orientation marked by the dark red pixel. White regions indicate angular orientations ($\theta$, $\phi$) not sampled by the docking program (i.e., below the scoring threshold).

To clarify the heatmap interpretation, Fig. 3 shows the docking-derived PNI map for serotransferrin (PDB ID: 6JAS) on a $SiO_2$ NP, with $\phi$ on the *x*-axis, $\theta$ on the *y*-axis, and color indicating the amplitude of the docking score. The strongest-scoring orientation appears as the dark red region. Each bin corresponds to a single representative docked complex. White regions do not imply forbidden orientations; rather, they indicate bins for which the docking protocol did not return a scored pose (e.g., below the sampling threshold).

Next, we constructed ($\theta$, $\phi$) heatmaps for both the docking and UAM datasets. Figure 4 shows the masked maps of the raw data using a 30° bin size. The two columns ("Docking" and "UAM") indicate the method used, and each row corresponds to a different protein.

To enable a direct, like-for-like comparison, we apply the same orientation mask to the corresponding UAM heatmaps.

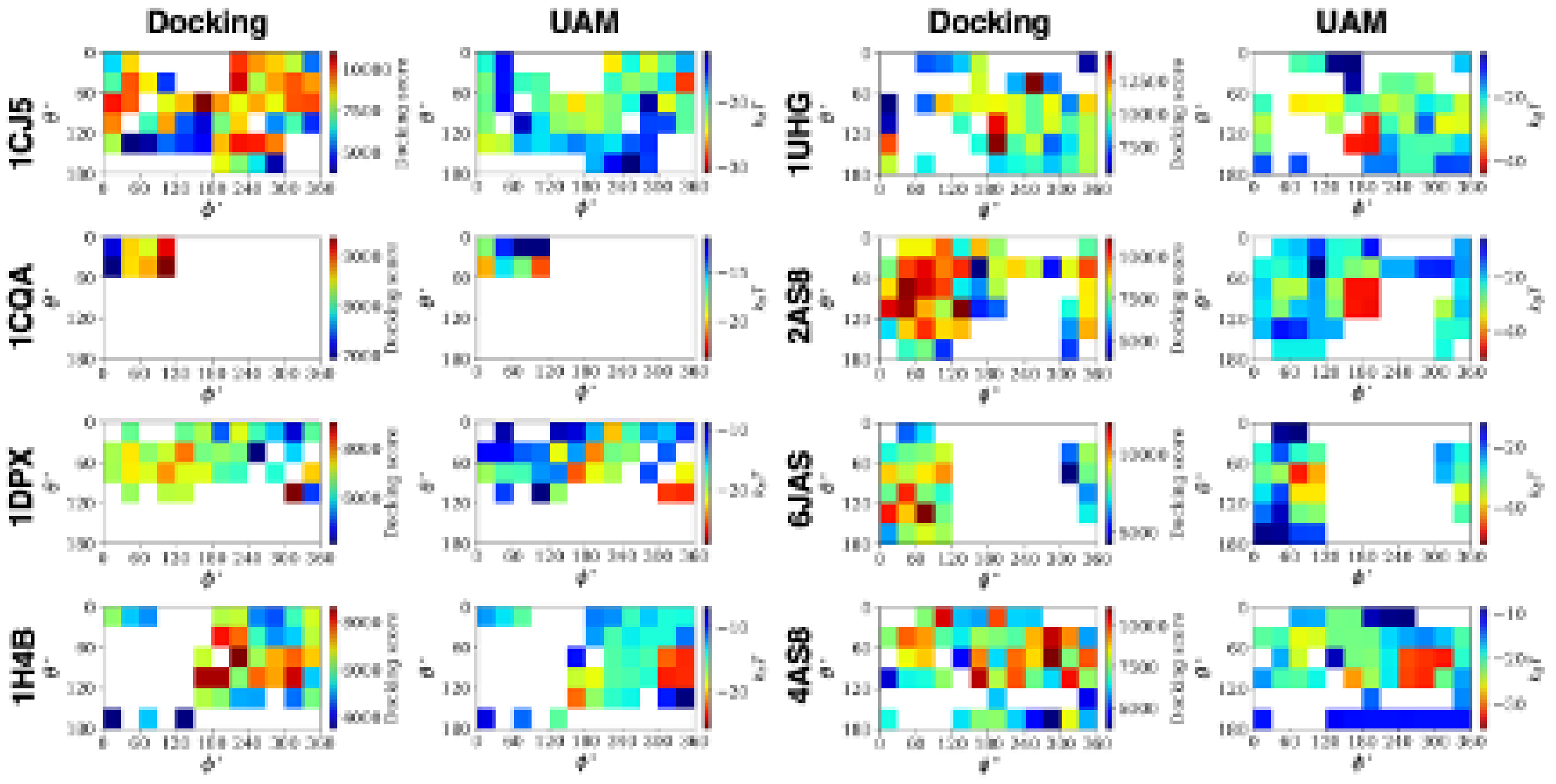

**Figure 4. Comparison of docking-based and UAM PNI maps.** For the eight birch pollen proteins (Fig. 1), PNI heatmaps computed using protein-NP docking and the united-atom model (UAM) are compared at a 30° ($\theta$, $\phi$) bin size. As in Fig. 3, white regions denote orientations not sampled by the docking program; the corresponding bins are masked in the UAM maps to enable direct comparison.

We note that both the docking and UAM approaches evaluated here assume a well-defined, folded protein conformation that remains relatively rigid and does not undergo substantial structural rearrangement upon binding to the nanoparticle (NP) surface. In cases where conformational changes do occur after adsorption, the protein reference frame defined above can still be maintained, provided that a unique structural alignment can be performed between the initial and NP-bound conformations. In practice, we expect that most proteins experience only modest conformational adjustments upon NP binding. Consequently, the rigid-protein approximation should not significantly affect the identification of dominant binding regions in either docking or UAM, thereby enabling a systematic comparison between the two methods.

Figure 5 presents line plots of Jensen–Shannon divergence (JSD) values as a function of bin size, illustrating the relationship between the docking heatmap and the Boltzmann-averaged adsorption energy from UAM for eight proteins. JSD quantifies the similarity between probability distributions, with values approaching zero indicating comparable probabilities and values approaching one reflecting distinct distributions.[39,40, 41,42,43] We note that the values shown in the PNI maps - both the docking scores and the UAM-derived quantities - were normalised prior to constructing their orientational distribution maps. This normalisation ensures that the Jensen–Shannon divergence (JSD) can meaningfully capture the similarity between the corresponding PNI landscapes, particularly with respect to the relative positions of their peaks and valleys.

For completeness, the Supplementary Material also includes the Earth Mover's Distance (EMD), which provides an alternative measure of distributional differences, where lower EMD values correspond to greater similarity.[44]

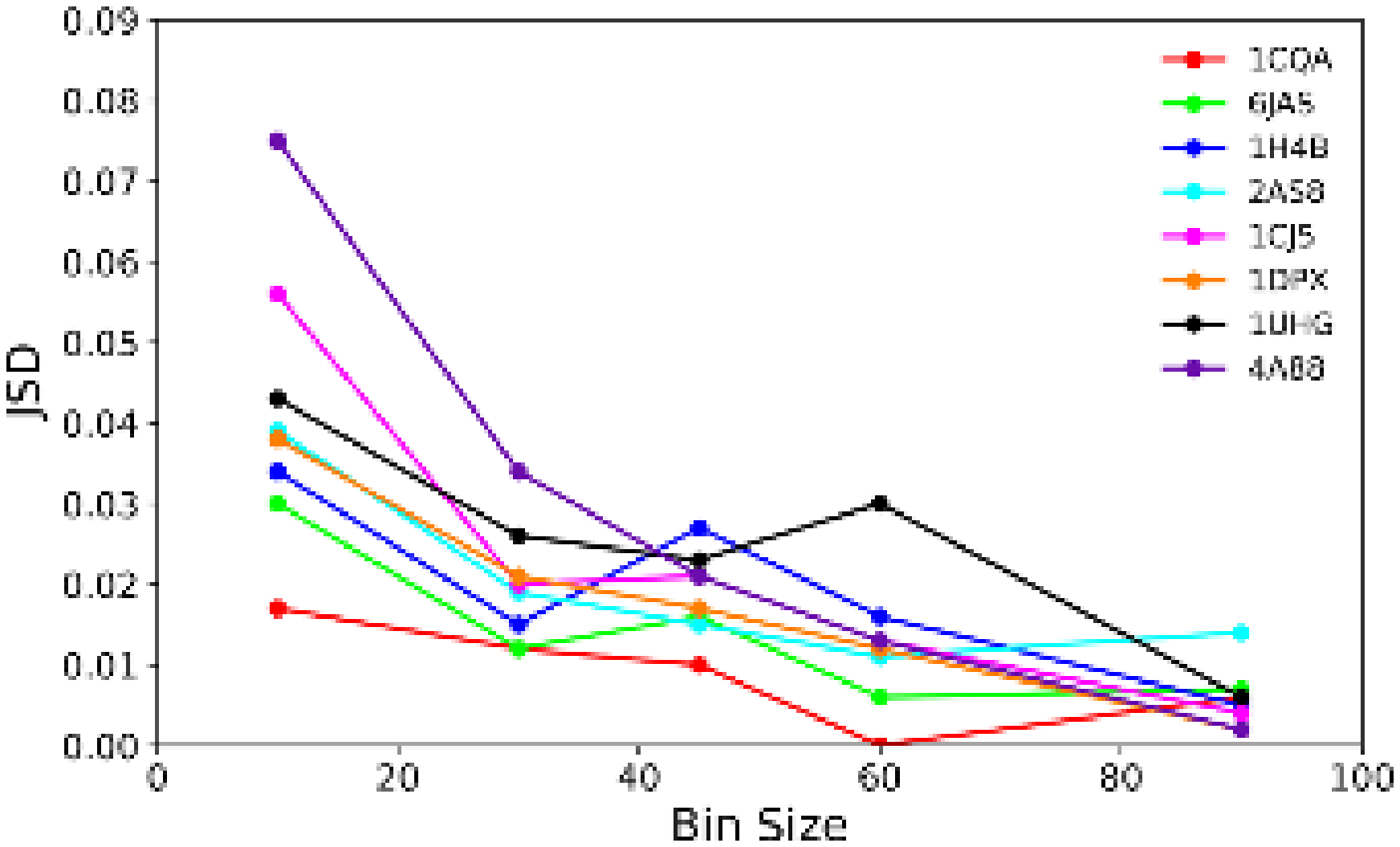

**Figure 5. Comparison of docking and UAM adsorption landscapes.** For the eight proteins (see Figs. 1 and 4), protein–NP docking score maps are compared with UAM Boltzmann-averaged adsorption energy maps as a function of the ($\theta$, $\phi$) bin size. Similarity between each pair of orientational distributions is quantified using the Jensen–Shannon divergence, where JSD = 0 indicates identical distributions and larger values (approaching 1) reflect increasing dissimilarity.

Across the eight proteins studied here (Fig. 1), we observe that smaller, more globular proteins tend to exhibit greater agreement in preferred binding modes between the docking- and UAM-derived PNI heatmaps.

To illustrate the corresponding adsorption geometries, we select the highest-scoring docking pose (defined by its $\phi$ and $\theta$ bin) and the strongest-binding UAM orientation and visualize the associated structures. Figure 6 shows this comparison for lysozyme (PDB ID: 1DPX), including the docking and UAM PNI heatmaps, the best-scoring docking pose (highlighted by the red region), and the corresponding representative complexes from each method.

**Figure 6. Docking and UAM PNI maps for lysozyme – $SiO_2$.** Orientation-resolved PNI heatmaps are shown for the lysozyme protein (PDB ID: 1DPX) interacting with a $SiO_2$ NP. The strongest-binding orientation is highlighted in dark red (arrows).

To sum up, motivated by the growing availability of experimental benchmarks and validated coarse-grained descriptions of PNI, we developed a docking-based workflow to construct orientation-resolved PNI heatmaps and to compare them directly with UAM adsorption energetics.[20,45] The approach was applied to eight birch pollen allergen proteins,[20] enabling systematic, model-to-model comparison of binding modes across a diverse set of sizes and folds. The PNI heatmaps are conceptually similar to maps used previously for quantifying residue-residue[46,47] and residue-backbone[48] interactions in proteins,[47-50] and protein-nanomaterial interactions.[33,36]

Protein orientations were parameterized using a protein-fixed spherical coordinate frame in which $(\theta, \phi)$ uniquely specify the relative pose, together with the center-of-mass separation $d_{COM}$. To ensure a fair comparison between methods, orientations not sampled by the docking protocol were masked and the same mask was applied to the UAM maps. Similarity between the resulting orientational distributions was quantified using JSD, which is well-suited for comparing probability-like landscapes. Across the proteins considered here, a 30° ($\theta$, $\phi$) angular bin size provided the most robust comparisons; finer binning tended to over-resolve the landscapes and reduced the stability of the metrics. Consistent with the visual trends in the PNI maps, smaller and more globular proteins generally show higher agreement between docking and UAM-predicted preferred orientations, as quantified by the corresponding JSD values (Fig. 5).

Differences between the docking-score and UAM energy heatmaps observed in Figs. 4 and 5 are expected, given the distinct algorithms and physical quantities used by these approaches. Nevertheless, although moderate discrepancies emerge at finer angular resolutions, the coarse-grained landscapes are largely consistent. This convergence indicates that both methods capture the dominant physical factors governing protein–NP binding. Taken together, these results support the validity of both approaches and reinforce the robustness of their predictions for preferred protein orientations.

Finally, we extracted representative strongest-binding orientations from both approaches to link features in the heatmaps to specific adsorption geometries. Taken together, the combined PNI–JSD framework provides a practical route for relating docking score landscapes to coarse-grained adsorption energetics at protein–NP interfaces, and for identifying when docking can serve as an efficient proxy versus when improved parameterization or sampling is needed.

Future work may extend this framework to systematically refine both coarse-grained PNI models and docking protocols. One promising direction is iterative optimization that reduces topological and amplitude discrepancies between PNI maps for targeted protein-NP systems, thereby improving consistency with experimental data - as more becomes available - and enhancing agreement across different theoretical approaches. Our study can have applications to improving accuracy of *in silico* modeling of NPIs and NP-protein corona structures with implications in a broad range of fields, from nanomedicine (e.g., designing NP-based drug delivery systems[8,9]) to nanoinformatics, nanotoxicology, and nanosafety.[4]

## AUTHOR DECLARATIONS

### Conflict of Interest

The authors have no conflicts to disclose.

## ACKNOWLEDGMENTS

We wish to thank the DJEI/DES/SFI/HEA Irish Centre for High-End Computing (ICHEC) and the Sonic HPC cluster (UCD Research IT) for the provision of computational facilities and support. VKA and N-VB acknowledge the financial support received from the Irish Research Council, and from the European Union’s Horizon 2020 research and innovation program “NanoinformaTIX” (H2020-NMBP-14-2018, grant 814426). VL and IR acknowledges funding from EU Horizon 2020 grant No. 814572 (NanoSolveIT) and Horizon Europe grant No. 101092741 (nanoPASS).

## SUPPLEMENTARY MATERIAL

**Supplemental material** consists of a PDF file including thirteen supplementary figures (Figs. S1 to S13). Supplemental Figures S1–S3 present PNI heatmaps comparing docking-based and UAM pollen protein–nanoparticle interaction maps at orientation bin sizes of 45° (S1), 60° (S2), and 90° (S3), respectively. Supplemental Figures S4–S6 report (i) Jensen–Shannon divergence (S4), (ii) Earth Movers Distance using Boltzmann-averaged UAM energies (S5), and (iii) Earth Movers Distance using arithmetic-averaged UAM energies (S6) for eight pollen proteins, comparing docking-based and UAM adsorption landscapes across orientation bin sizes and quantifying similarity between the corresponding orientational distributions. Supplemental Figures S7–S13 present orientation-resolved adsorption maps for the remaining seven pollen-related proteins interacting with a $SiO_2$ nanoparticle, showing PNI heatmaps and indicating the strongest-binding orientation, consistent with the lysozyme–NP results shown in the main text (Figure 6).

# Supplemental Material:

# Orientation-Dependent Protein Binding at Nanoparticle Interfaces

*Vigneshwari Karunakaran Annapoorani*[1,2], *Ian Rouse*[1,2],

*Vladimir Lobaskin*[1,2], *Nicolae-Viorel Buchete*[1,2] *

[1]School of Physics, University College Dublin, Dublin 4, Belfield, Ireland

[2]Institute for Discovery, University College Dublin, Dublin 4, Belfield, Ireland

* Correspondence: buchete@ucd.ie

May 05th, 2026

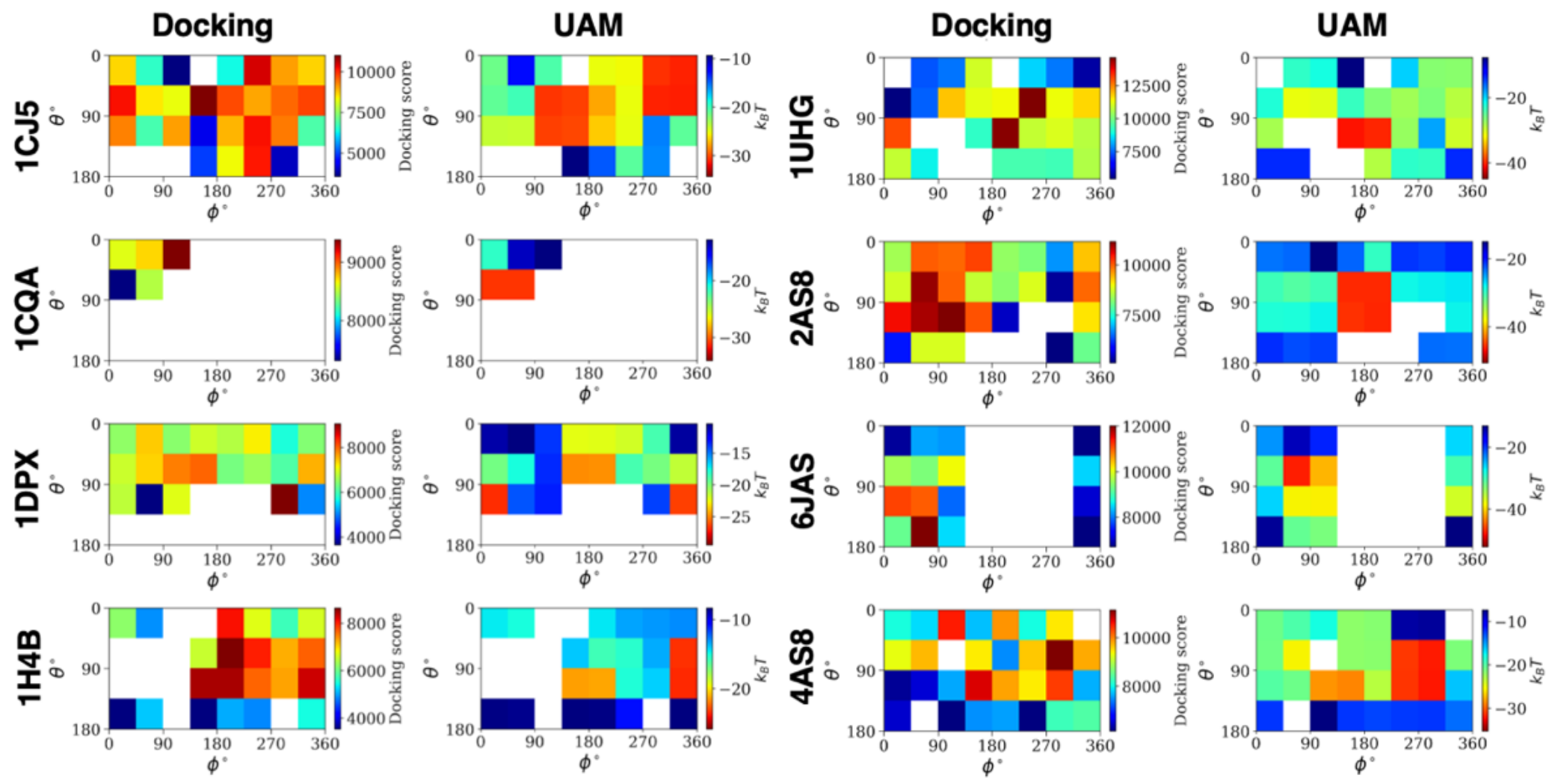

**Figure S1. Docking vs UAM PNI maps: raw data, 45° bin size.** For the eight birch pollen proteins (see Fig. 1), PNI heatmaps computed with protein–NP docking (PatchDock) and the united-atom model (UAM) are compared at a 45° ($\theta$, $\phi$) bin size. White regions indicate orientations not scored by docking; the corresponding bins are also masked (white) in the UAM maps for direct comparison.

**Figure S1:** Heatmaps comparing docking-based and UAM pollen protein–nanoparticle interaction maps at 45° orientation bins, with unscored docking orientations shown as masked regions for consistent comparison.

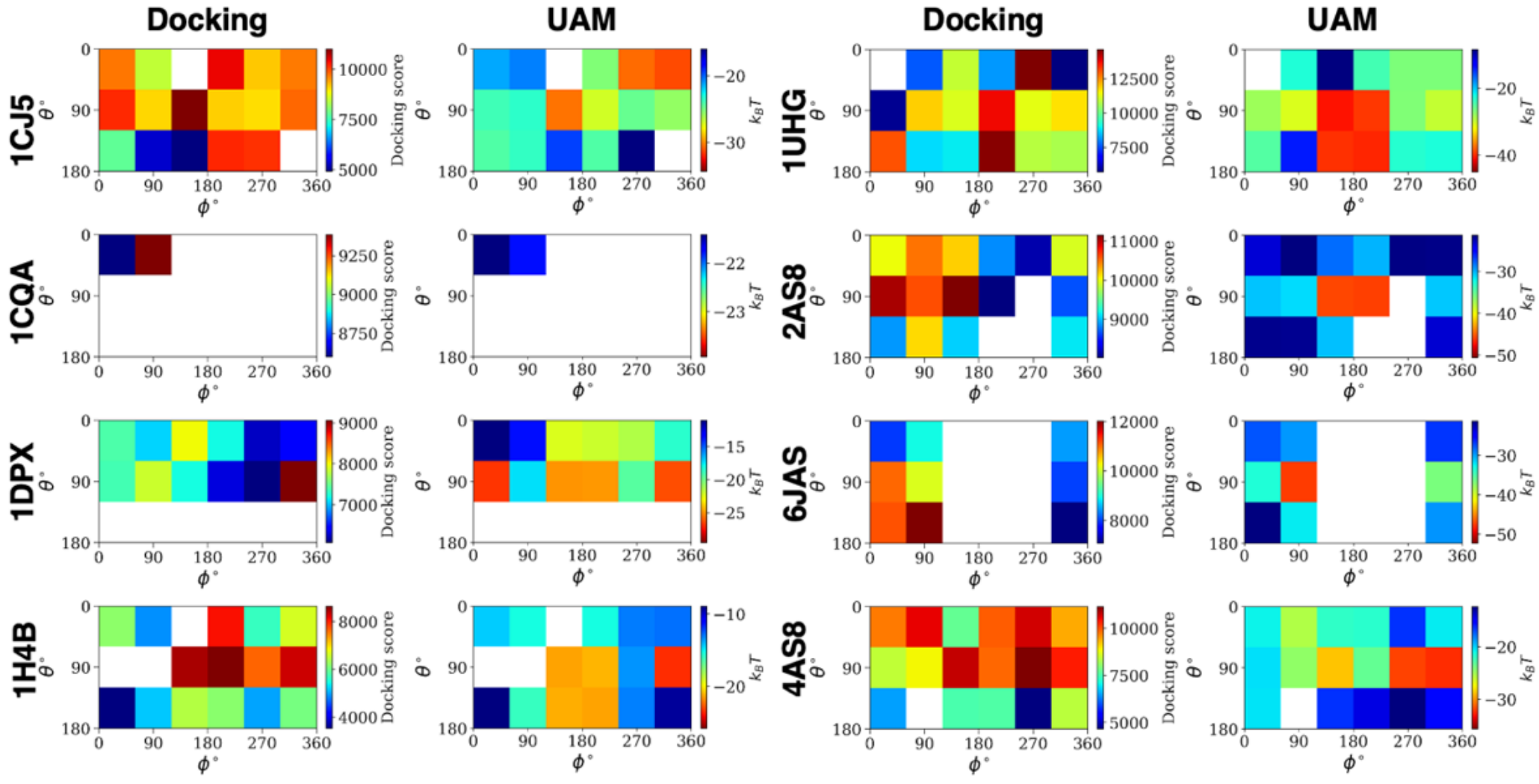

**Figure S2. Docking vs UAM PNI maps: raw data, 60° bin size.** For the eight birch pollen proteins (see Fig. 1), PNI heatmaps computed with protein–NP docking (PatchDock) and the united-atom model (UAM) are compared at a 60° ($\theta$, $\phi$) bin size. White regions indicate orientations not scored by docking; the corresponding bins are also masked (white) in the UAM maps for direct comparison.

**Figure S2:** Heatmaps comparing docking-based and UAM pollen protein–nanoparticle interaction maps at 60° orientation bins, with unscored docking orientations shown as masked regions to allow direct comparison across methods.

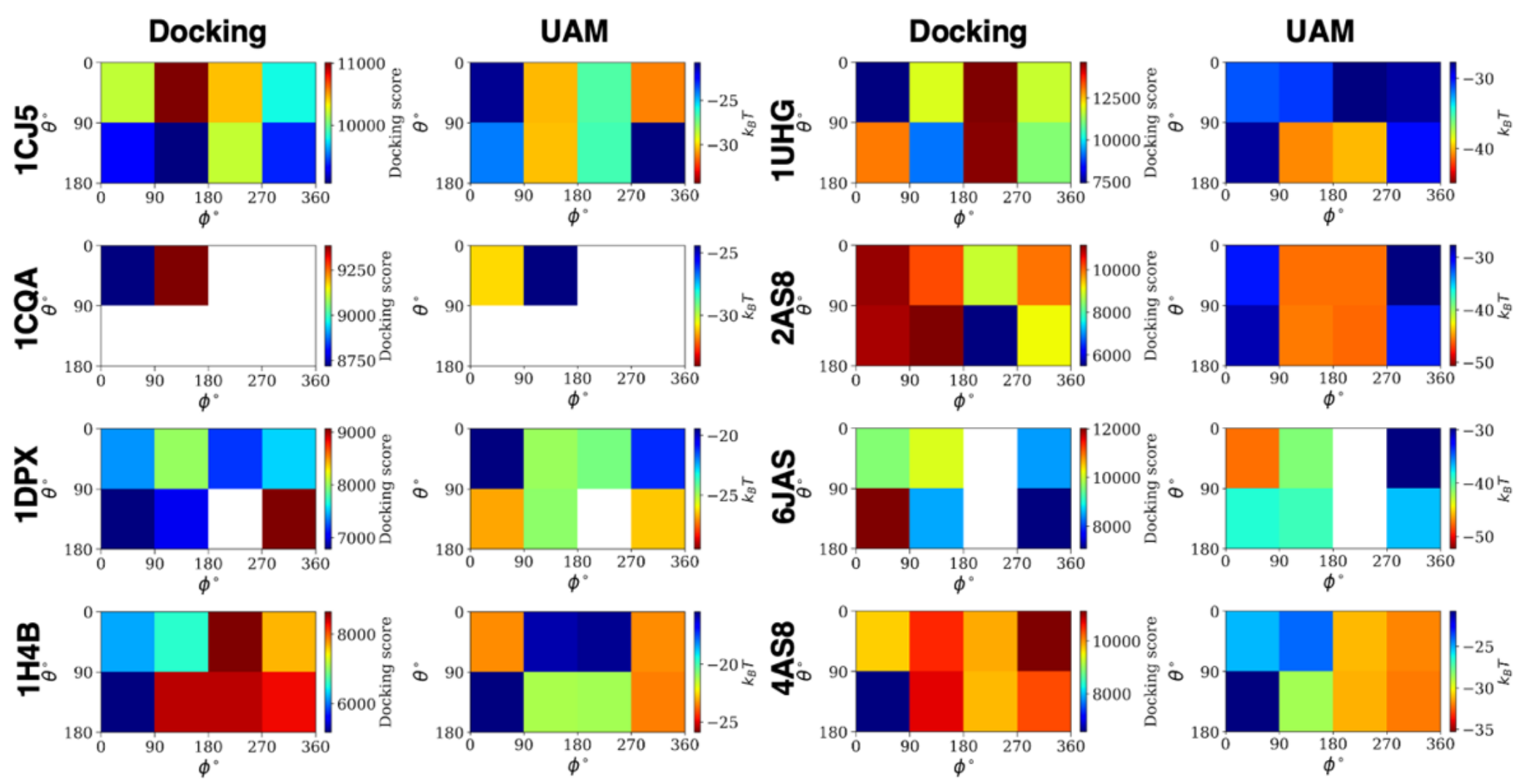

**Figure S3. Docking vs UAM PNI maps: raw data, 90° bin size.** For the eight birch pollen proteins (see Fig. 1), PNI heatmaps computed with protein–NP docking (PatchDock) and the united-atom model (UAM) are compared at a 90° ($\theta$, $\phi$) bin size. White regions indicate orientations not scored by docking; the corresponding bins are also masked (white) in the UAM maps for direct comparison.

**Figure S3:** PNI maps comparing docking-based and UAM pollen protein–nanoparticle interaction maps at 90° orientation bins, with unscored docking orientations masked to maintain consistent comparison across methods.

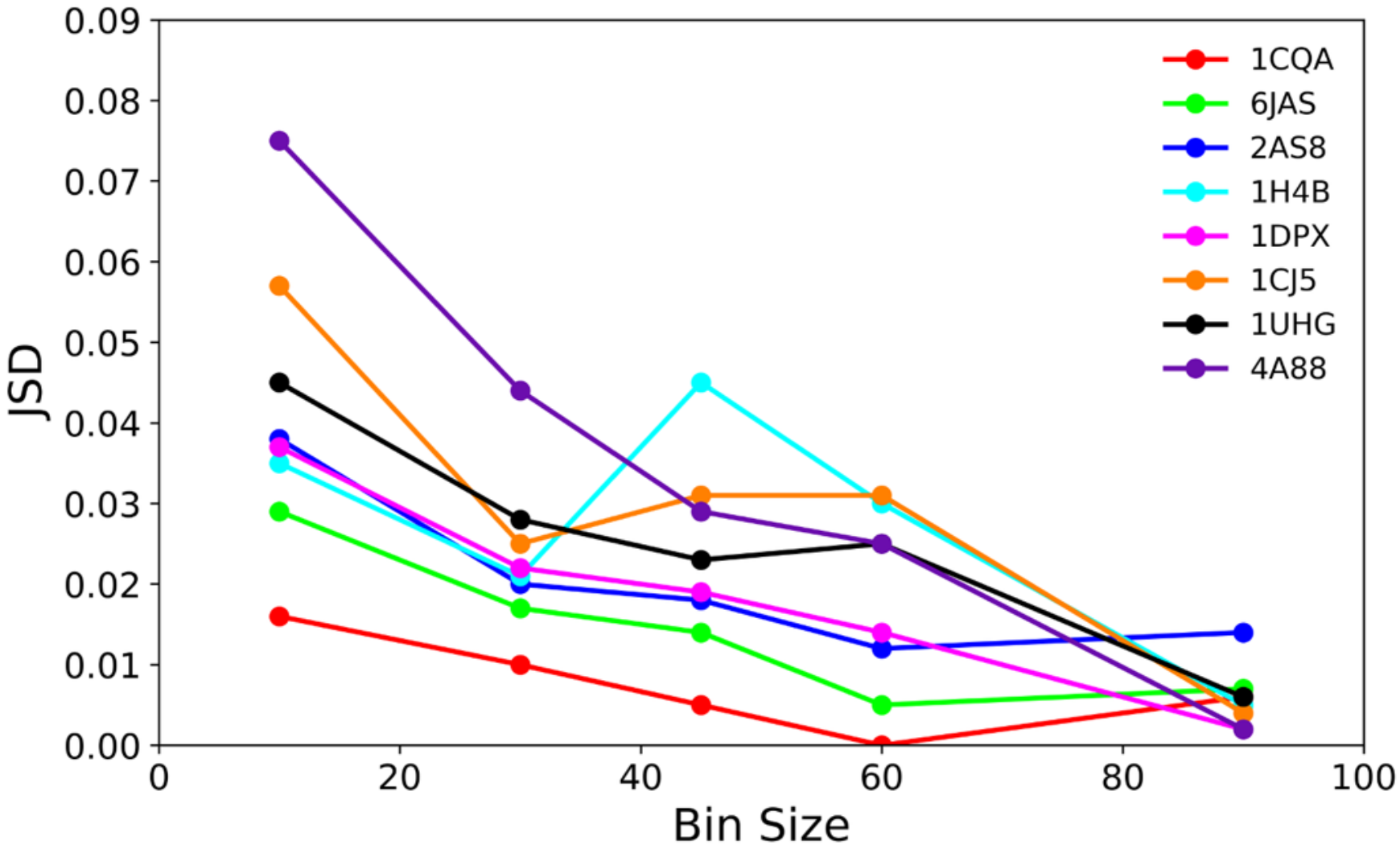

**Figure S4. Comparison of docking and UAM adsorption landscapes: Jensen–Shannon divergence (JSD), raw data (see Figs. S1-S3).** For the eight proteins, we compare protein–NP docking score maps with UAM raw data adsorption energy maps as a function of the ($\theta$, $\phi$) bin size. Similarity between each pair of orientational distributions is quantified by JSD values: JSD = 0 indicates identical distributions, whereas larger values (approaching 1) indicate increasing dissimilarity.

**Figure S4:** Jensen–Shannon divergence values comparing docking-based and UAM adsorption landscapes for eight pollen proteins across different orientation bin sizes, quantifying similarity between the corresponding orientational distributions.

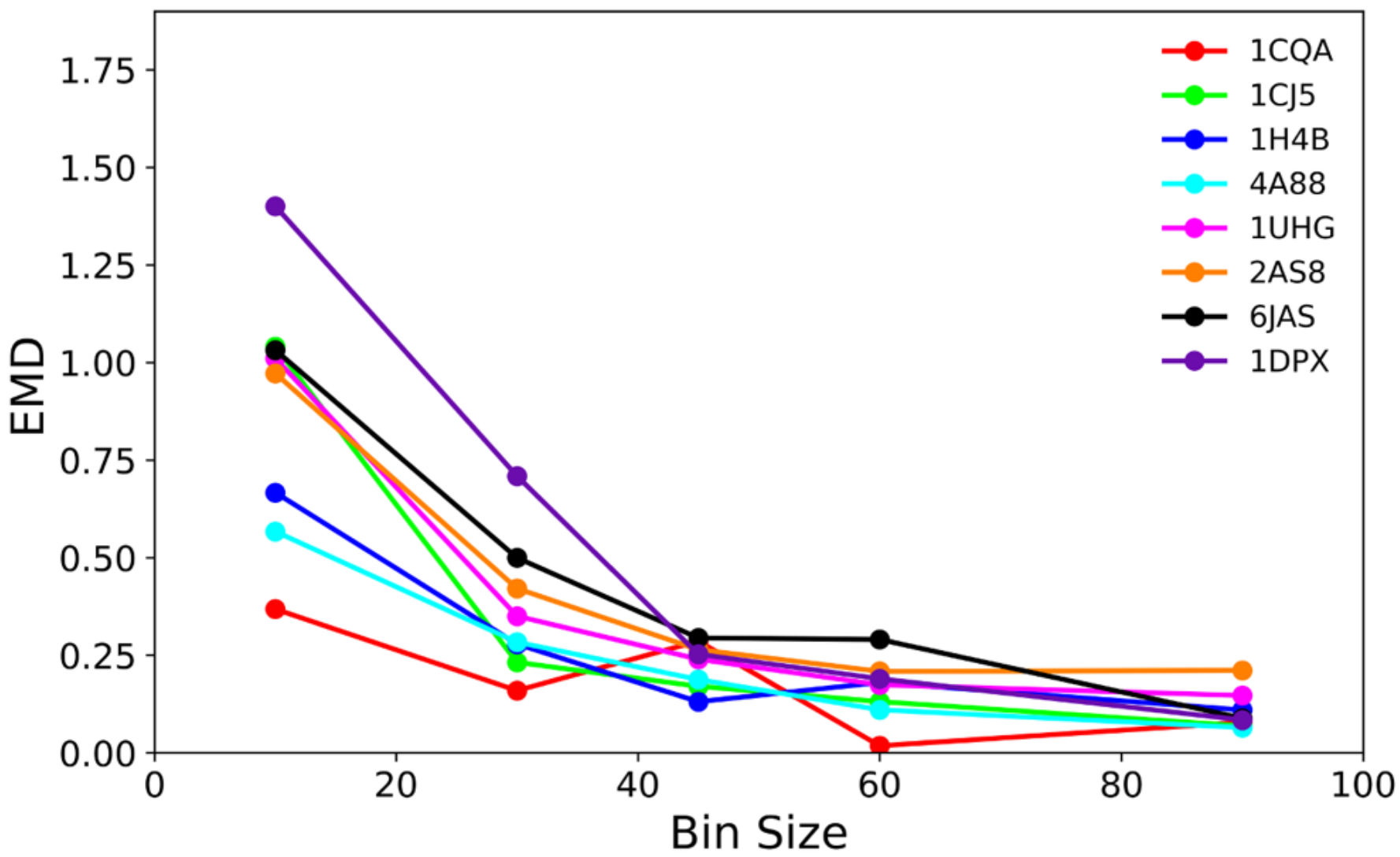

**Figure S5. Comparison of docking and UAM adsorption landscapes: Earth Movers Distance (EMD), Boltzmann averages (see Figs. S1-S3).** For the eight proteins, we compare protein–NP docking score maps with UAM Boltzmann-averaged adsorption energy maps as a function of the ($\theta$, $\phi$) bin size. Similarity between each pair of orientational distributions is quantified by EMD values: EMD = 0 indicates identical distributions, whereas larger values indicate increasing dissimilarity.

**Figure S5:** Earth Movers Distance values comparing docking-based and UAM Boltzmann-averaged adsorption landscapes for eight pollen proteins across orientation bin sizes, quantifying similarity between the corresponding orientational distributions.

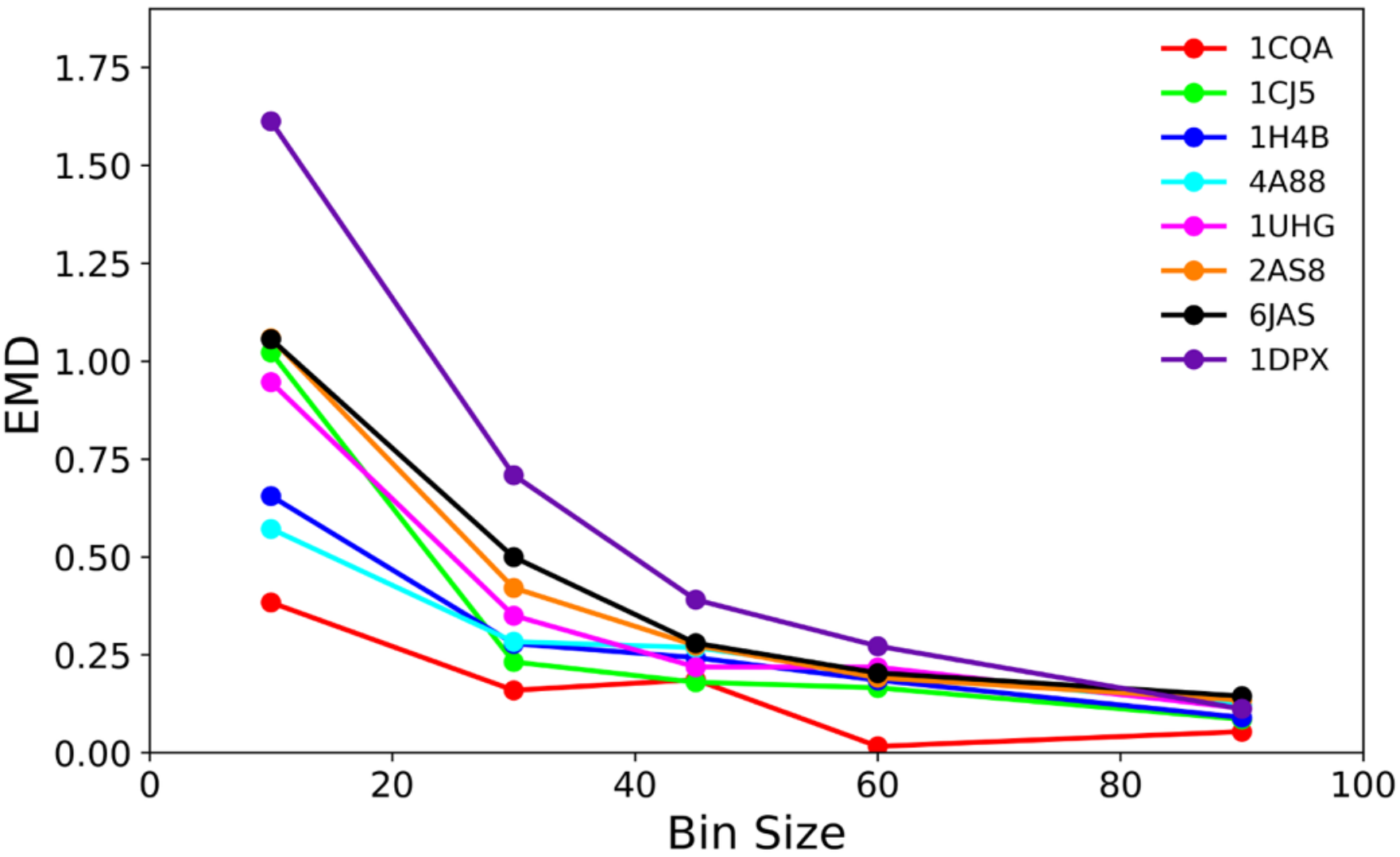

**Figure S6. Comparison of docking and UAM adsorption landscapes: Earth Movers Distance (EMD), arithmetic averages (see Figs. S1-S3).** For the eight proteins, we compare protein–NP docking score maps with UAM arithmetically-averaged adsorption energy maps as a function of the ($\theta$, $\phi$) bin size. Similarity between each pair of orientational distributions is quantified by EMD values: EMD = 0 indicates identical distributions, whereas larger values indicate increasing dissimilarity.

**Figure S6:** Earth Movers Distance values comparing docking-based and UAM arithmetic-average adsorption landscapes for eight pollen proteins across orientation bin sizes, measuring similarity between the corresponding orientational distributions.

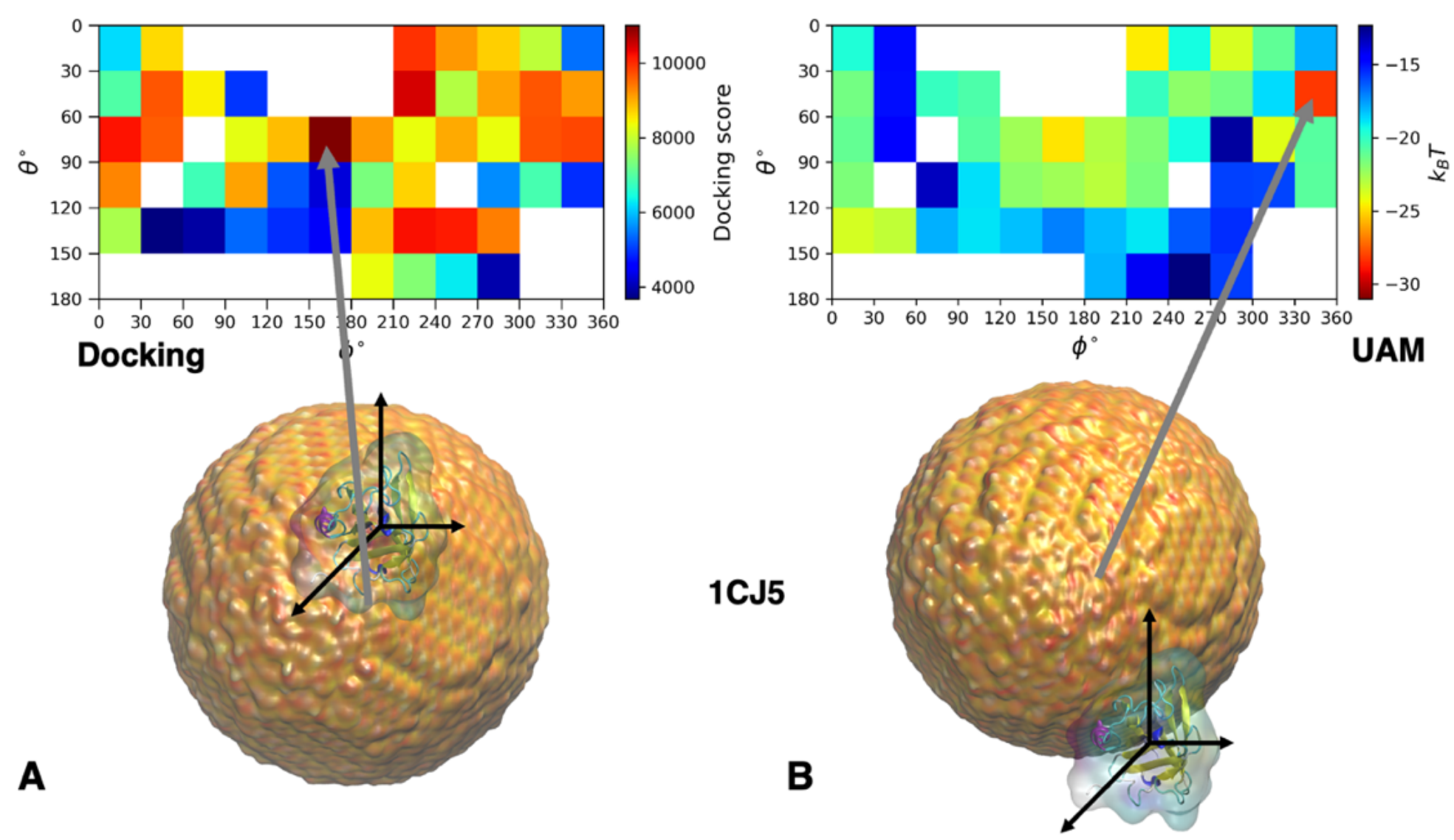

**Figure S7. Docking and UAM PNI maps for β-lactoglobulin A – $SiO_2$.** Orientation-resolved PNI heatmaps are shown for the β-lactoglobulin A (PDB ID: 1CJ5) interacting with a $SiO_2$ nanoparticle. The strongest-binding orientation is highlighted in dark red (arrows).

**Figure S7:** Orientation-resolved adsorption maps for β-lactoglobulin A interacting with a $SiO_2$ nanoparticle, showing PNI heatmaps and marking the strongest-binding orientation.

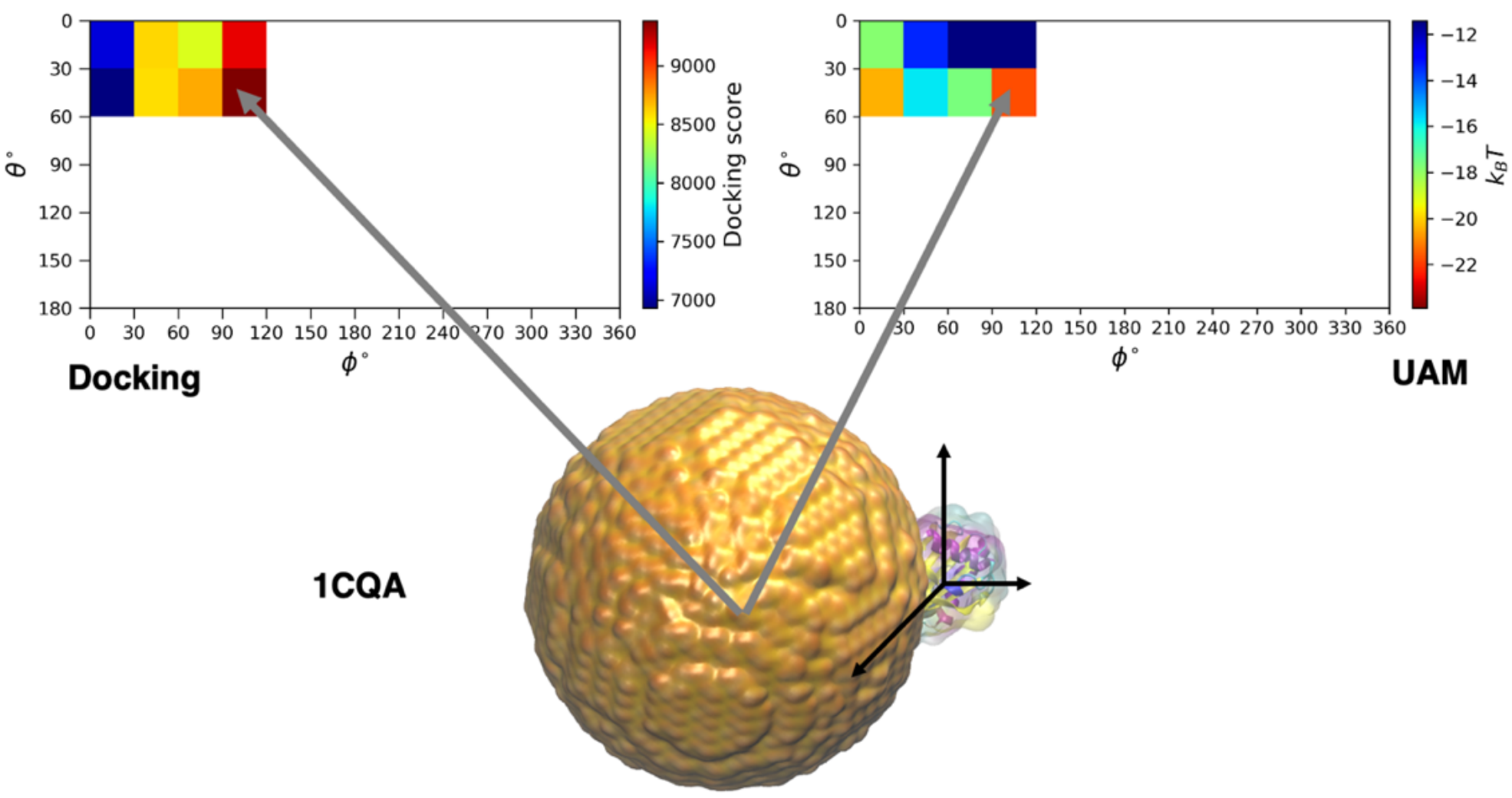

**Figure S8. Docking and UAM PNI maps for Bet v 2 – $SiO_2$.** Orientation-resolved PNI heatmaps are shown for the Bet v 2 protein (1CQA) interacting with a $SiO_2$ nanoparticle. The strongest-binding orientation is highlighted in dark red (arrows).

**Figure S8:** Orientation-resolved adsorption maps for Bet v 2 interacting with a $SiO_2$ nanoparticle, showing PNI heatmaps and marking the strongest-binding orientation.

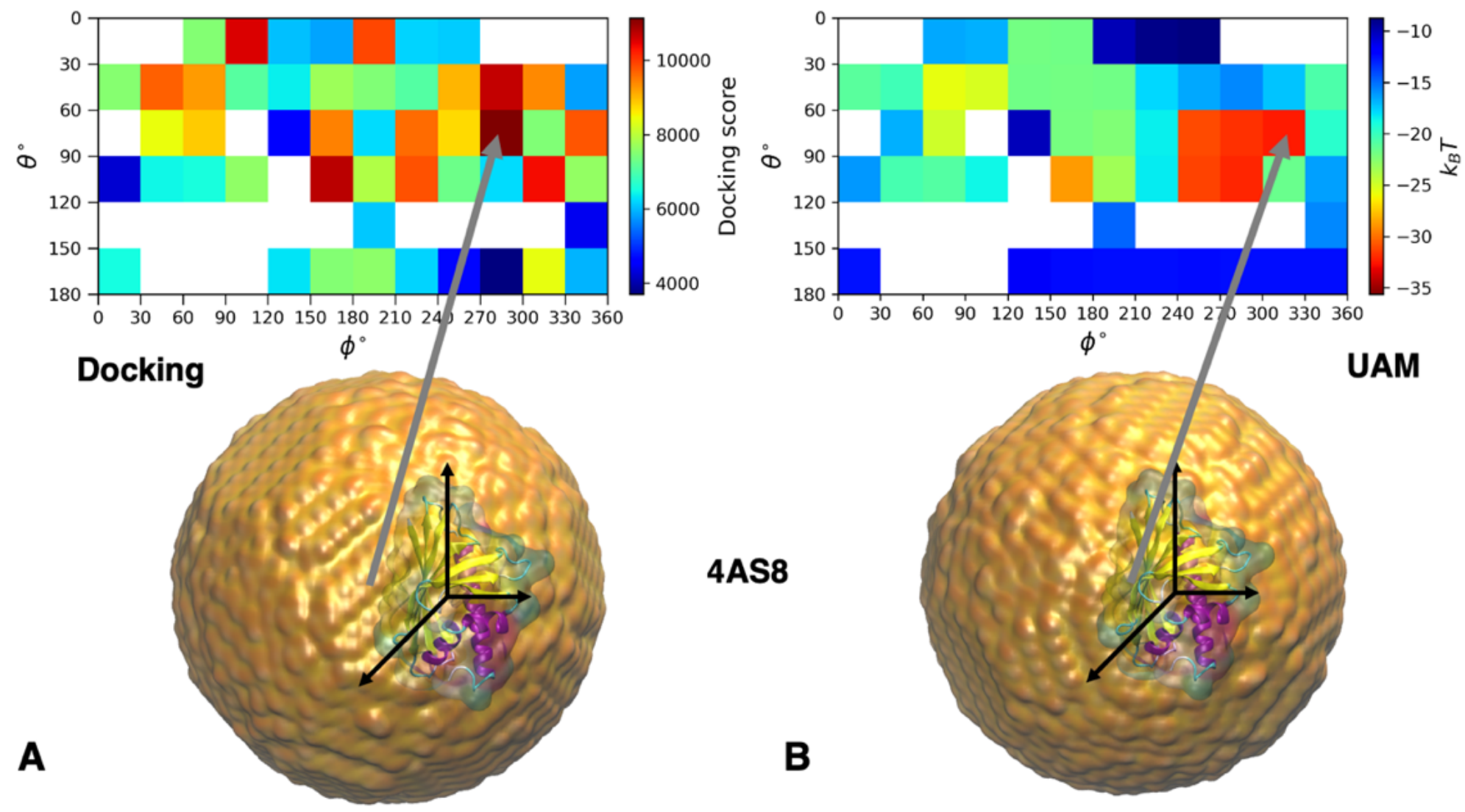

**Figure S9. Docking and UAM PNI maps for Bet v 1 – $SiO_2$.** Orientation-resolved PNI heatmaps are shown for the Bet v 1 protein (4A88) interacting with a $SiO_2$ nanoparticle. The strongest-binding orientation is highlighted in dark red (arrows).

**Figure S9**: Orientation-resolved adsorption maps for Bet v 1 interacting with a $SiO_2$ nanoparticle, showing PNI heatmaps and marking the strongest-binding orientation.

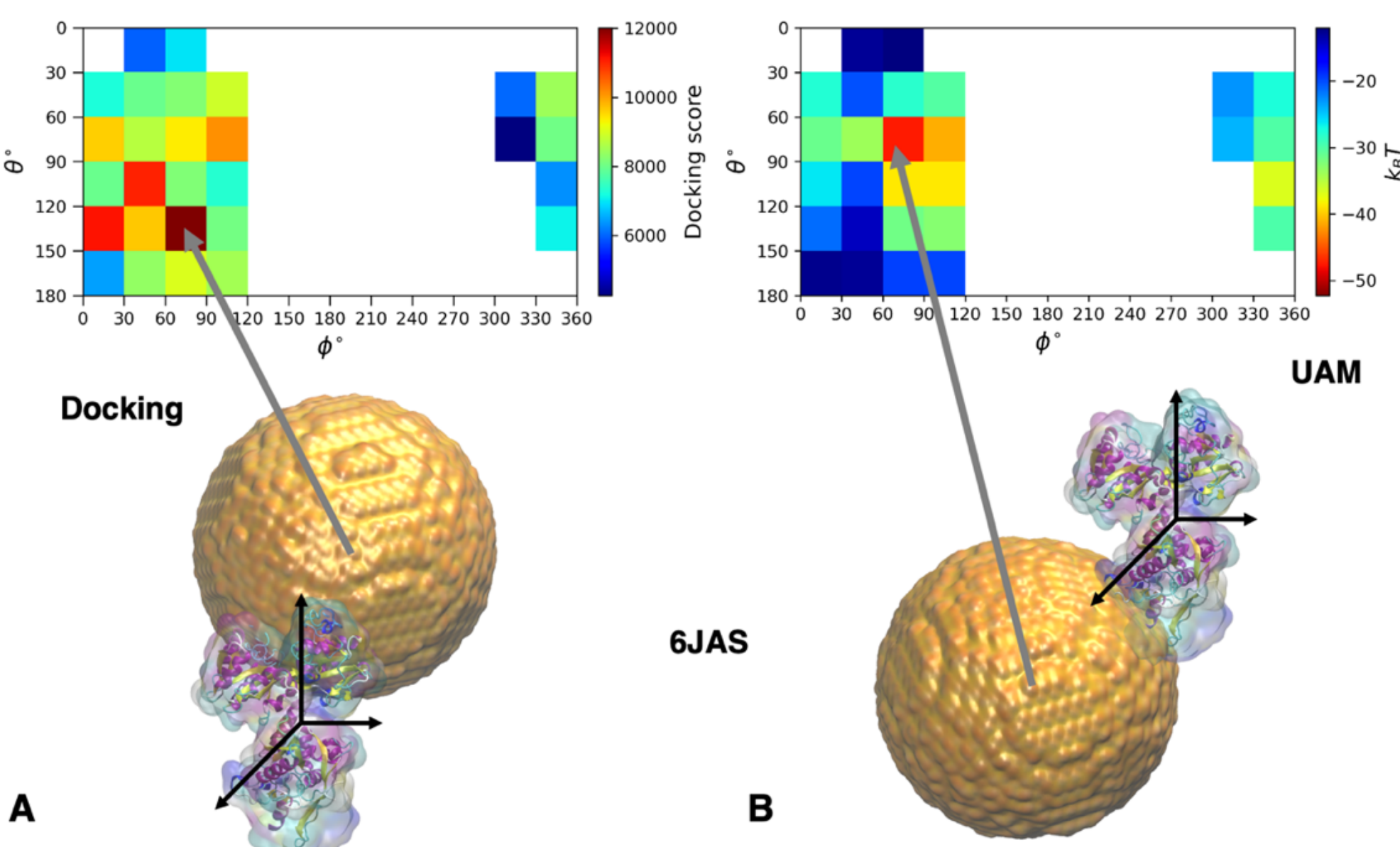

**Figure S10. Docking and UAM PNI maps for serotransferrin – $SiO_2$.** Orientation-resolved PNI heatmaps are shown for serotransferrin (6JAS) interacting with a $SiO_2$ nanoparticle. The strongest-binding orientation is highlighted in dark red (arrows).

**Figure S10**: Orientation-resolved adsorption maps for serotransferrin interacting with a $SiO_2$ nanoparticle, showing PNI heatmaps and marking the strongest-binding orientation.

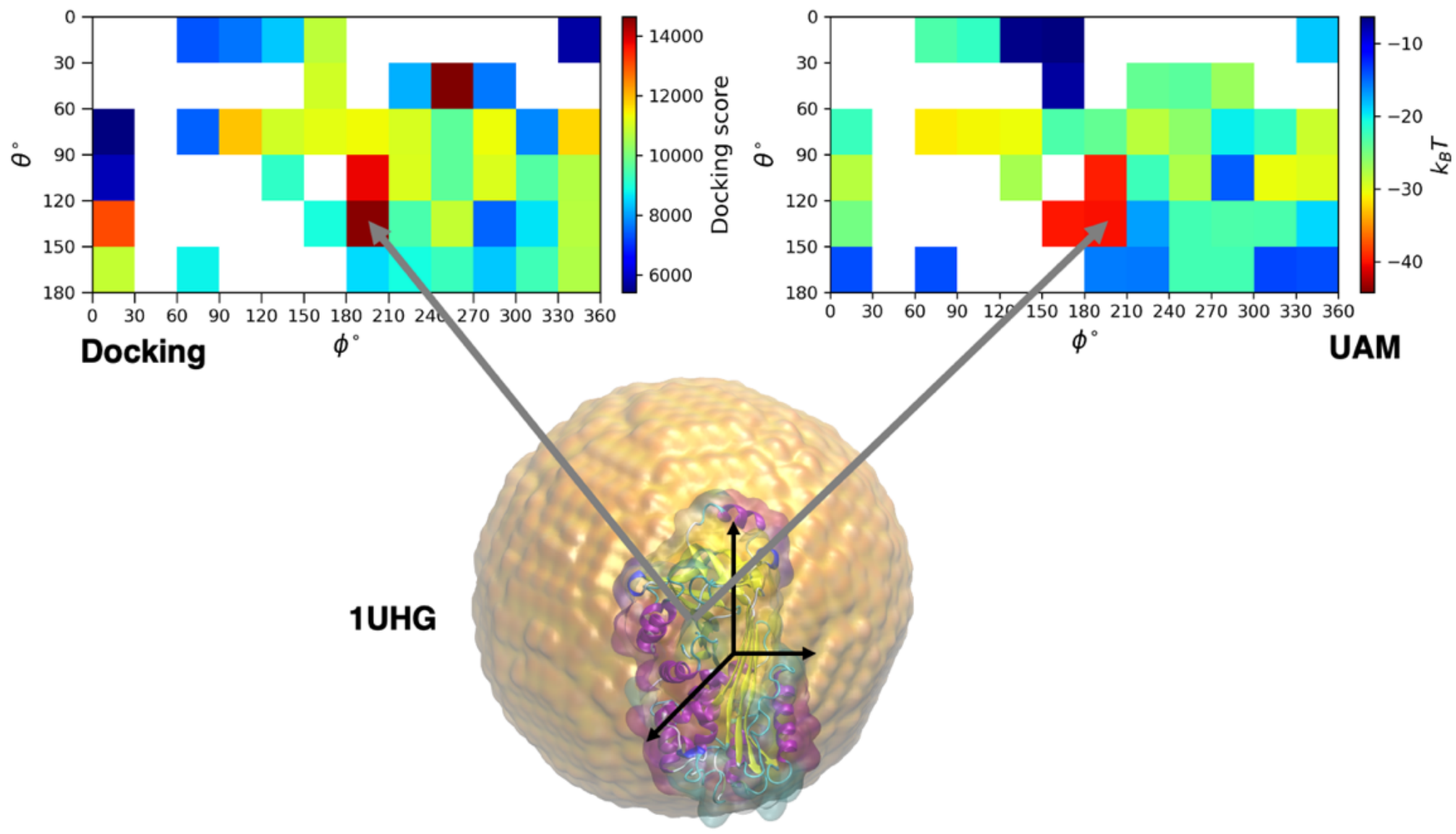

**Figure S11. Docking and UAM PNI maps for ovalbumin – $SiO_2$.** Orientation-resolved PNI heatmaps are shown for ovalbumin (1UHG) interacting with a $SiO_2$ nanoparticle. The strongest-binding orientation is highlighted in dark red (arrows).

**Figure S11**: Orientation-resolved adsorption maps for ovalbumin interacting with a $SiO_2$ nanoparticle, showing PNI heatmaps and marking the strongest-binding orientation.

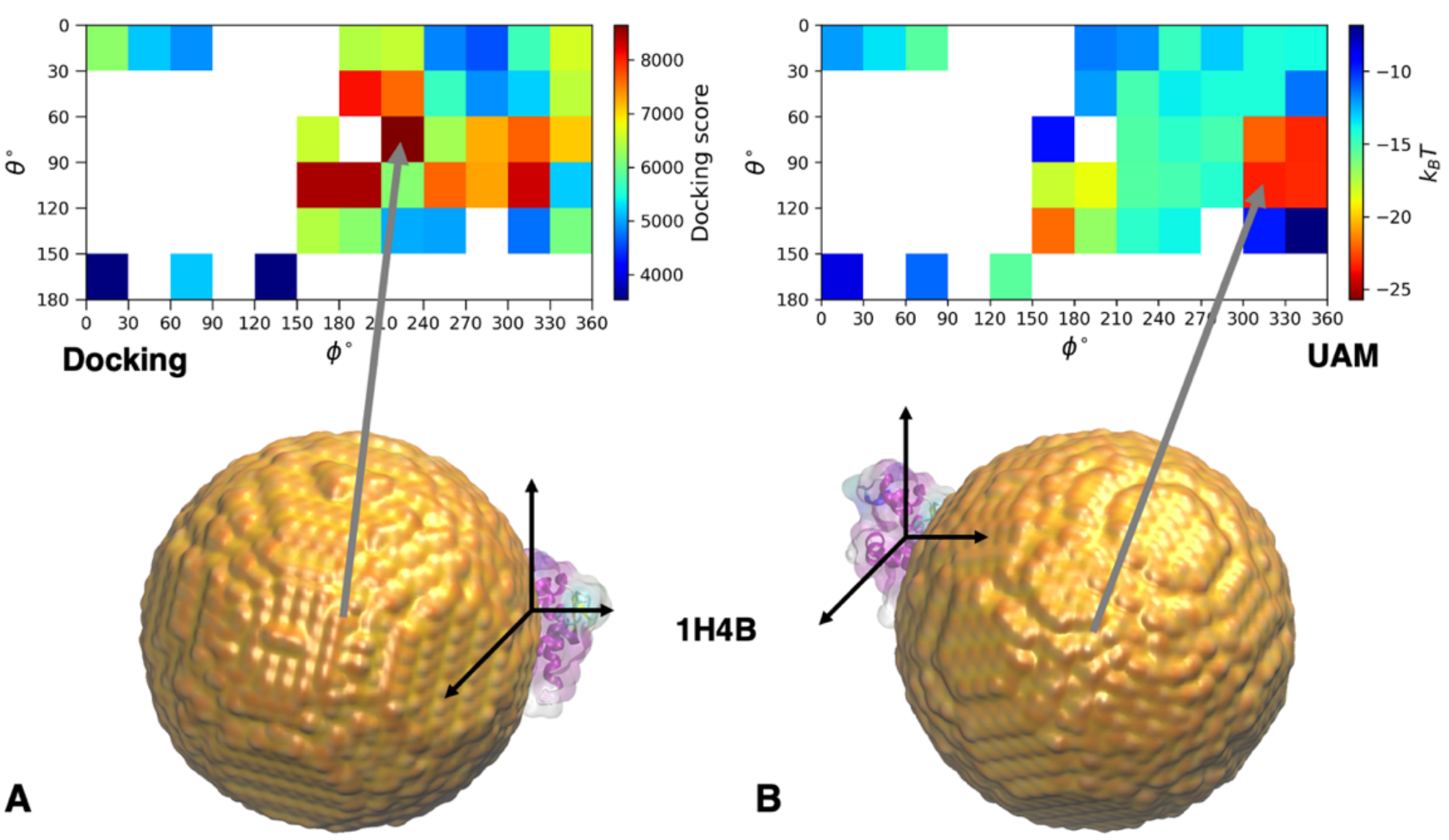

**Figure S12. Docking and UAM PNI maps for Bet v 4 – $SiO_2$.** Orientation-resolved PNI heatmaps are shown for the Bet v 4 protein (1H4B) interacting with a $SiO_2$ nanoparticle. The strongest-binding orientation is highlighted in dark red (arrows).

**Figure S12**: Orientation-resolved adsorption maps for Bet v 4 protein interacting with a $SiO_2$ nanoparticle, showing PNI heatmaps and marking the strongest-binding orientation.

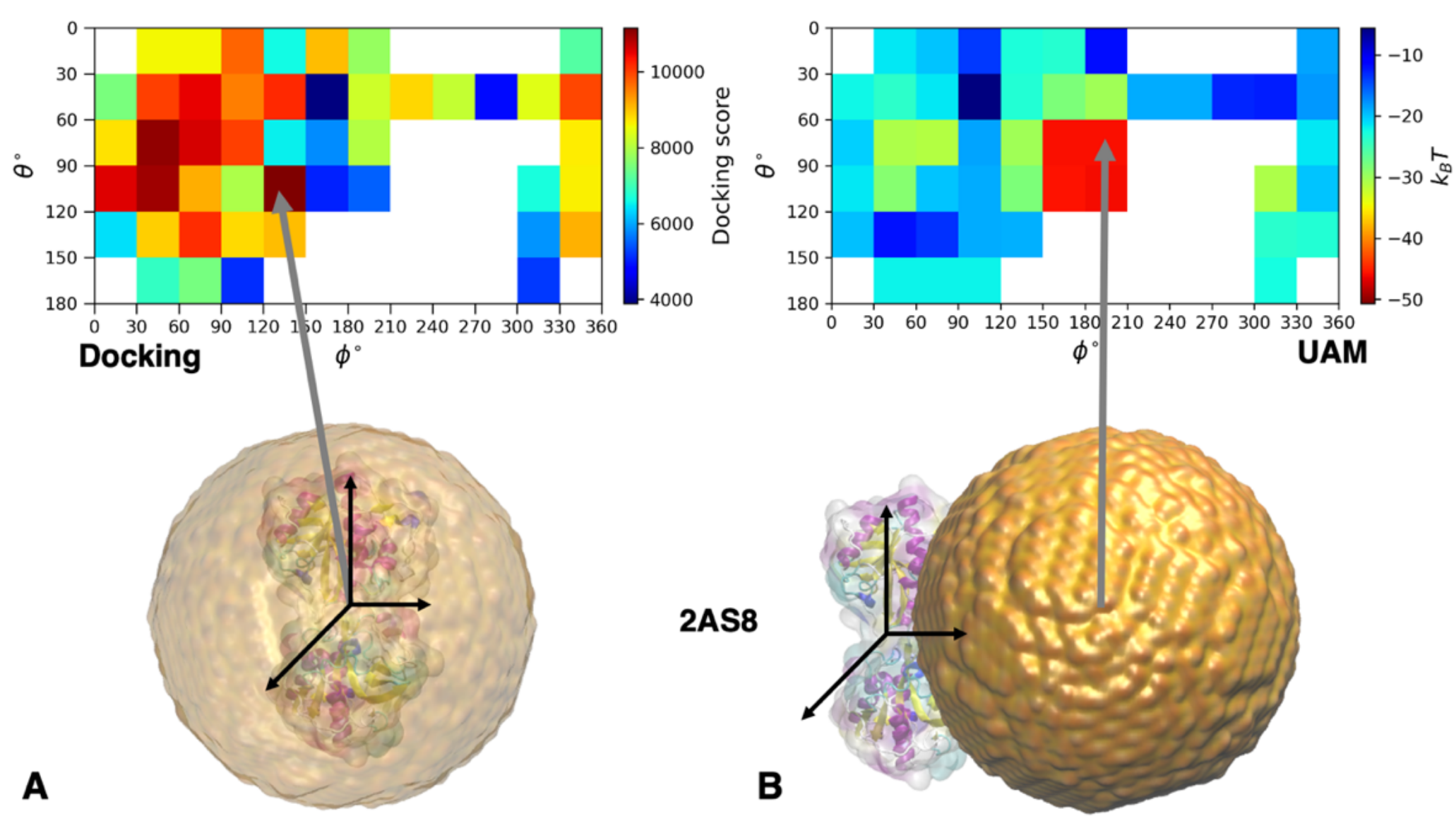

**Figure S13. Docking and UAM PNI maps for Bet v 6 – $SiO_2$.** Orientation-resolved PNI heatmaps are shown for the Bet v 6 protein (2AS8) interacting with a $SiO_2$ nanoparticle. The strongest-binding orientation is highlighted in dark red (arrows).

**Figure S13**: Orientation-resolved adsorption maps for Bet v 6 protein interacting with a $SiO_2$ nanoparticle, showing PNI heatmaps and marking the strongest-binding orientation.